\documentclass[twocolumn,tighten]{aastex63}

\usepackage{times,natbib,graphicx,amsmath,multirow,xspace}
\usepackage{xcolor}
\usepackage{lineno}
\newcommand{\nustar}{NuSTAR\xspace}
\newcommand{\swift}{Swift\xspace}
\newcommand{\chandra}{Chandra\xspace}
\newcommand{\nicer}{NICER\xspace}

\newcommand{\integral}{INTEGRAL\xspace}
\newcommand{\rxte}{RXTE\xspace}

\newcommand{\xmm}{XMM-Newton\xspace}
\newcommand{\lc}{light curve}
\newcommand{\ms}{$M_{\odot}$\xspace}
\newcommand{\mdot}{$\dot{\mathrm{m}}$}
\newcommand{\fluxcgs}{ergs~cm$^{-2}$~s$^{-1}$\xspace}
\newcommand{\lumcgs}{ergs~s$^{-1}$\xspace}
\newcommand{\rin}{$R_{\rm in}$\xspace}
\newcommand{\rg}{$R_{g}$\xspace}
\newcommand{\risco}{$R_{\mathrm{ISCO}}$\xspace}

\newcommand{\xillver}{{\sc xillver}\xspace}
\newcommand{\xillverco}{{\sc xillverCO}\xspace}
\newcommand{\source}{\mbox{4U~1543$-$624}\xspace}

\shorttitle{NICER-NuSTAR View of 4U 1543$-$624}
\shortauthors{Ludlam et al.}

\begin{document}

\title{Simultaneous \nicer and \nustar Observations of the Ultra-compact X-ray Binary 4U 1543$-$624}

\correspondingauthor{R. M. Ludlam}
\email{rmludlam@caltech.edu}

\author[0000-0002-8961-939X]{R.~M.~Ludlam}\thanks{NASA Einstein Fellow}
\affiliation{Cahill Center for Astronomy and Astrophysics, California Institute of Technology, Pasadena, CA 91125, USA}

\author[0000-0002-3850-6651]{A.~D.~Jaodand}
\affiliation{Cahill Center for Astronomy and Astrophysics, California Institute of Technology, Pasadena, CA 91125, USA}

\author[0000-0003-3828-2448]{J.~A.~Garc\'{i}a}
\affiliation{Cahill Center for Astronomy and Astrophysics, California Institute of Technology, Pasadena, CA 91125, USA}
\affiliation{Remeis Observatory \& ECAP, Universit\"{a}t Erlangen-N\"{u}rnberg, Sternwartstr. 7, D-96049, Bamberg, Germany}

\author{N.~Degenaar}
\affiliation{Anton Pannekoek Institute for Astronomy, University of Amsterdam, Pastbus 94249, 1090 GE Amsterdam, The Netherlands}

\author{J.~A.~Tomsick}
\affiliation{Space Sciences Laboratory, 7 Gauss Way, University of California, Berkeley, CA 94720-7450, USA}

\author[0000-0002-8294-9281]{E.~M.~Cackett}
\affiliation{Department of Physics \& Astronomy, Wayne State University, 666 West Hancock Street, Detroit, MI 48201, USA}

\author{A.~C.~Fabian}
\affiliation{Institute of Astronomy, Madingley Road, Cambridge CB3 0HA, UK}

\author{P.~Gandhi}
\affiliation{Department of Physics and Astronomy, University of Southampton, Highfield, Southampton, SO17 1BJ}

\author{D.~J.~K.~Buisson}
\affiliation{Department of Physics and Astronomy, University of Southampton, Highfield, Southampton, SO17 1BJ}

\author{A.~W.~Shaw}
\affiliation{Department of Physics, University of Nevada, Reno, NV 89557, USA}

\author{D.~Chakrabarty}
\affiliation{MIT Kavli Institute for Astrophysics and Space Research, Massachusetts Institute of Technology, Cambridge, MA 02139, USA}

\begin{abstract}
We present the first joint \nustar and \nicer observations of the ultra-compact X-ray binary (UCXB) \source obtained in 2020 April. The source was at a luminosity of $L_{0.5-50\ \mathrm{keV}} = 4.9 (D/7\ \mathrm{kpc})^{2}\times10^{36}$~\lumcgs and showed evidence of reflected emission in the form of an \ion{O}{8} line, Fe~K line, and Compton hump within the spectrum.
We used a full reflection model, known as \xillverco, that is tailored for the atypical abundances found in UCXBs, to account for the reflected emission. We tested the emission radii of the O and Fe line components and conclude that they originate from a common disk radius in the innermost region of the accretion disk (\rin$ \leq1.07$~\risco).  Assuming that the compact accretor is a neutron star (NS) and the position of the inner disk is the Alfv\'{e}n radius, we placed an upper limit on the magnetic field strength to be $B\leq0.7(D/7\ \mathrm {kpc})\times10^{8}$~G at the poles. Given the lack of pulsations detected and position of \rin, it was likely that a boundary layer region had formed between the NS surface and inner edge of the accretion disk with an extent of 1.2~km. This implies a maximum radius of the neutron star accretor of $R_{\mathrm{NS}}\leq 12.1$~km when assuming a canonical NS mass of 1.4~\ms.

\end{abstract}


\section{Introduction} \label{sec:intro}
Ultra-compact X-ray binaries (UCXBs) are a subclass of low-mass X-ray binaries (LMXBs) with short orbital periods of $\lesssim90$ minutes.
The tight orbit of these systems means the compact object, either a neutron star (NS) or black hole (BH), is accreting via Roche-lobe overflow from a degenerate stellar companion, such as a white dwarf or He star \citep{nelson86, savonije86}. UCXBs are strong, persistent gravitational wave sources for future missions, such as NASA/ESA's LISA, that are sensitive in the sub-mHz regime \citep{nelemans10}. 

The accretion disks in these systems differ from those of typical LMXBs since they are almost devoid of hydrogen while overabundant in oxygen, carbon, and/or neon \citep{nelemans03}. When accretion disks are externally illuminated by hard X-rays originating from close to the compact object, the photons are reprocessed and re-emitted as a series of atomic features superimposed onto a `reflected' continuum. These features are then broadened due to Doppler, special, and general relativistic effects in this region \citep{fabian00}. The strength of these effects depend on the proximity to the compact object, therefore, these reflection features can be used to infer fundamental properties of the compact object, as well as the accretion disk itself (e.g., \citealt{miller07, cackett08, cackett09b, cackett10, papitto09, disalvo09, disalvo15, miller13, ludlam17a}). 

In a typical accretion disk composed of solar abundance material, the Fe K line at $6.4-6.97$~keV is the most prominent feature.  However, in an UCXB, \ion{O}{8} ($\sim0.65$ keV) becomes dominant over Fe~K \citep{ballantyne02}. It was previously thought that Fe emission should not be present in these systems since most of the ionizing radiation within the disk would be absorbed by the lower-Z atomic elements \citep{koliopanos13}. However, this was revealed not to be the case via \xmm\ and \chandra\ observations of the UCXBs 4U~1543$-$624 and 4U~0614+091 \citep{madej11,madej14}.
Indeed other UCXBs observed with \xmm have since shown evidence of Fe emission lines (e.g., 4U~1728$-$34, 4U~1820$-$30, 4U~1916$-$05: \citealt{koliopanos20a}), although some detections are marginal (e.g., MAXI J0911$-$655: \citealt{sanna17}).  
Additionally, \nustar\ has observed reflection features in 4U~0614+091 \citep{ludlam19a} and the recently classified UCXB IGR J17062-6143 \citep{degenaar17, eijnden18,strohmayer18}.
The predicted absence of Fe emission in these systems was based on models which assume a cold, neutral disk  and, therefore, any ionizing photons have a higher probability of being absorbed by the overabundant O atoms rather than Fe (see Fig. 1 in \citealt{koliopanos13}). Yet, the observational evidence of the Fe~K line in UCXBs implies that the disk is hot and being illuminated in a similar manner to other accreting LMXBs \citep{madej14}. 

\source is an UCXB with an orbital period of \hbox{$18.2\pm0.1$} minutes \citep{wang04, wang15} located at a distance between \hbox{$D\sim1.4-11.5$~kpc} \citep{wang04,bailerjones18,serino18}. The nature of the compact object in \source is uncertain, but very likely a NS from a tentative association with a Type-1 X-ray burst seen by MAXI \citep{serino18} and its radio--X-ray behavior \citep{ludlam17d,ludlam19b,tetarenko18}. The degenerate companion in this system is a C/O or O/Ne white dwarf due to the absence of hydrogen and helium lines coupled with emission from carbon and oxygen in the optical spectrum \citep{nelemans03}.

\begin{figure}[b!]
\begin{center}
\includegraphics[width=0.48\textwidth,trim=5 0 0 0,clip]{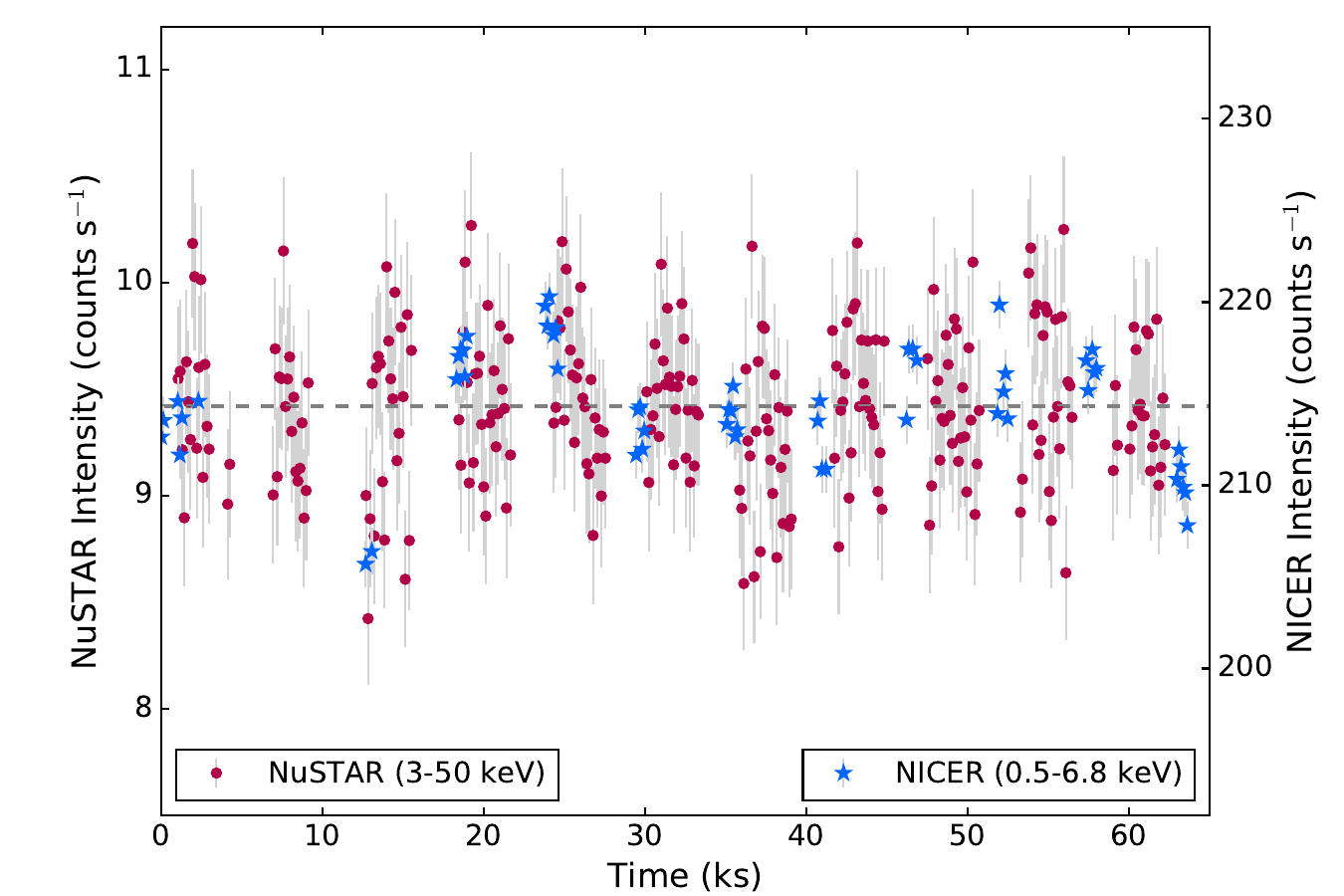}
\caption{Light-curve for the \nustar/FPMA (circles) and \nicer\ (stars) observations of \source binned to 128~s.  The grey dashed line indicates the average count rate for both \nustar\ and \nicer. 
The time elapsed is from the start of the \nicer observation on 2020-04-19 at 07:12:55UT. The source exhibits $\lesssim10\%$ variability over the course of the observation. Only one FPM is shown for clarity.}
\label{fig:lc}
\end{center}
\end{figure}

As mentioned previously, the X-ray spectrum of \source has shown a broad O {\sc viii} Ly$\alpha$ emission feature at $\sim0.7$ keV in conjuction with Fe K emission \citep{juett03,madej11}.  \citet{madej14} presented an X-ray spectral analysis of \source and 4U~0614+091 using a preliminary version of a new reflection model, \xillverco, that was tailored to accommodate the atypical elemental abundances in UCXBs. This model mimics the negligible H and He abundances in the disk by setting the abundance of metals to 10 times solar abundance and allowing for variable abundance of C and O. Though this only had a limited number of grid points (i.e., large steps between parameter values), spectral modeling using this initial \xillver\ grid on \source\ indicated an inner disk radius $<7.4$ \rg\ (where $R_{g}=GM/c^{2}$) and an inclination of  $i\sim65^{\circ}$ \citep{madej14}.

\begin{figure} [!t]
\begin{center}
\includegraphics[width=0.48\textwidth]{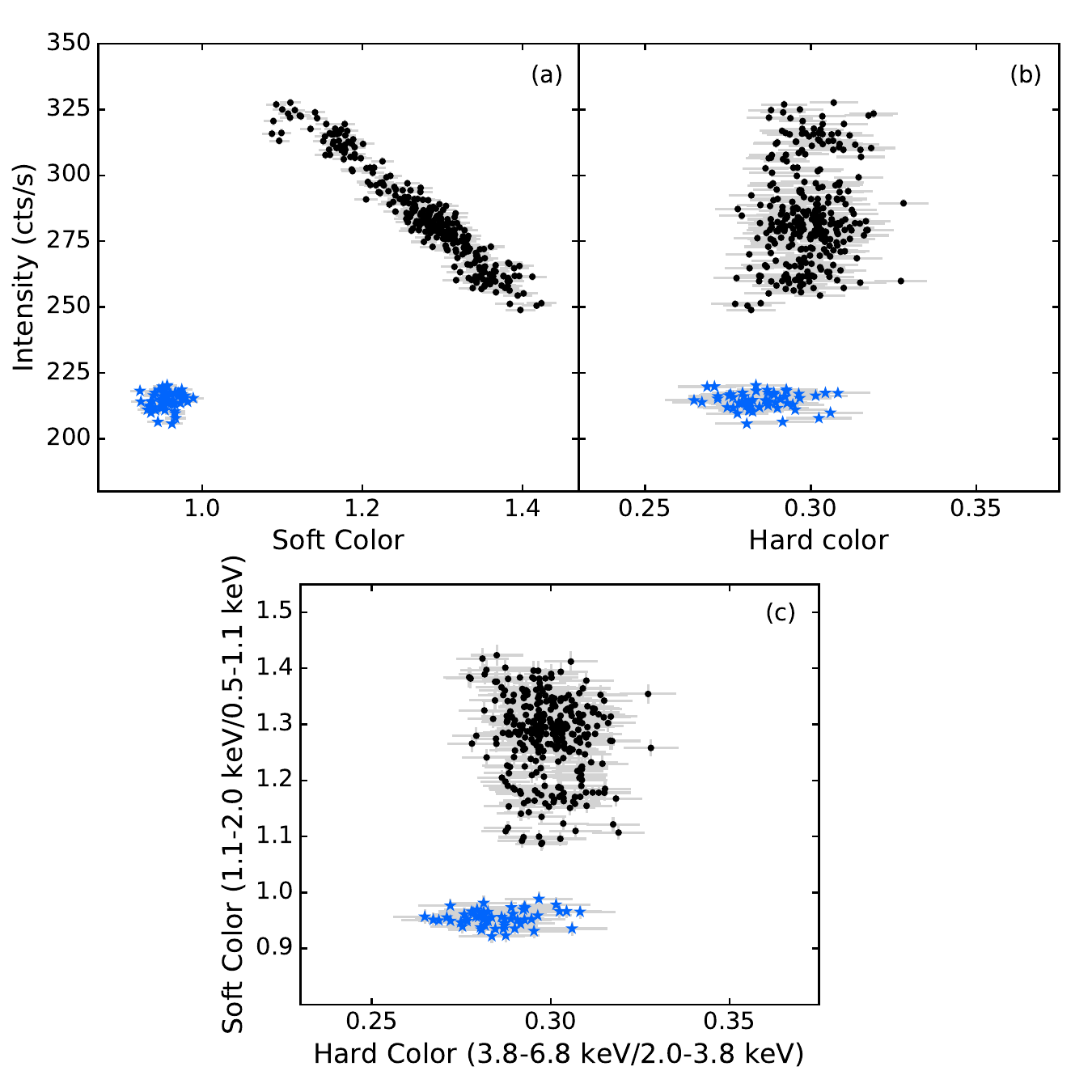}
\caption{A comparison of the \nicer observations of \source during the 2017 outburst (black circles) reported in \citet{ludlam19b} to the observations obtained in 2020 (blue stars) for: a) the soft color versus the source intensity in the $0.5-6.8$ keV band, b) the hard color versus intensity, and c) the soft color versus the hard color. The new observations probe different regions on these planes.}
\label{fig:HID}
\end{center}
\end{figure}

More recently, \citet{ludlam19b} reported on \nicer monitoring of \source over a $\sim10$~day period in 2017 August while the source underwent a period of enhanced accretion activity with supplemental observations by \swift, \integral, and ATCA. The monitoring of this event allowed for tracking of changes in the accretion disk in this system. There was an increase in the strength of the thermal component at the lowest energies as the accretion disk moved closer to the NS (from \rin$>60$~\rg to \rin$<8$~\rg at peak intensity: \citealt{ludlam19b}). There was also a clear change in the shape and strength of the emission lines as well, however, this analysis simply used {\sc diskline} to model the emission lines from Fe and O rather than a full reflection spectrum framework leading to uncertainties regarding a common emission radius for these two features.

We present simultaneous observations of \source with \nicer and \nustar from 2020 April. This is the first time that \nustar has observed the source.
The goal of these observations is to analyze the reflection spectrum in this system with a full reflection model to determine if the O and Fe components originate from similar disk radii and ionization.
The combined passband of \nicer and \nustar are ideal for revealing the presence of reflected emission while pinning down the continuum \citep{ludlam20, wang20}. We present the observations and data reduction in \S \ref{sec:data}, our analysis in \S \ref{sec:results}, and discuss the results in \S \ref{sec:discussion}.

\section{Observations and Data Reduction} \label{sec:data}

\nicer observed \source twice during the span of the contemporaneous \nustar observation. The first observation, ObsID 3604010101, began at 07:09:05 UT on 2020 April 19 for an exposure of 9.1~ks. The second observation, ObsID 3604010102, began at 00:33:20 UT on 2020 April 20 for 863~s.  
The \nicer observations were reduced using {\sc nicerdas} 2020-04-23\_V007a. Data were re-calibrated with the latest calibration files available in CALDB release 20200722 through implementation of the {\sc nicerl2} command. Good time intervals (GTIs) were generated using {\sc nimaketime} to select events that occurred when the particle background was low (KP~$<$~5 and COR\_SAX~$>$~4) and avoiding times of extreme optical light loading (SUN\_ANGLE~$>$~60 and FPM\_UNDERONLY\_COUNT~$<$~200)\footnote{See \citet{bogdanov19} regarding information on the \nicer\ screening flags.}. Using {\sc niextract-events}, the GTIs were applied to the data. The resulting event files were loaded into {\sc xselect} to extract a combined spectrum and \lc s in various energy bands. Background spectra were generated using the nibackgen3C50v6\footnote{https://heasarc.gsfc.nasa.gov/docs/nicer/tools/nicer\_bkg\_est\_tools.html} tool (R. Remillard, in prep.) for each cleaned and ufa (calibrated but unfiltered) event file pair based on instrument proxies to account for the observing conditions at the time. These were then combined into a single background spectrum that was weighted by the duration of each cleaned event file using {\sc mathpha}. We use the standard public RMF and the on-axis average ARF in CALDB v.20200722 when modeling the \nicer spectrum.

\begin{figure} 
\begin{center}
\includegraphics[angle=270,width=0.47\textwidth,trim=50 20 30 60,clip]{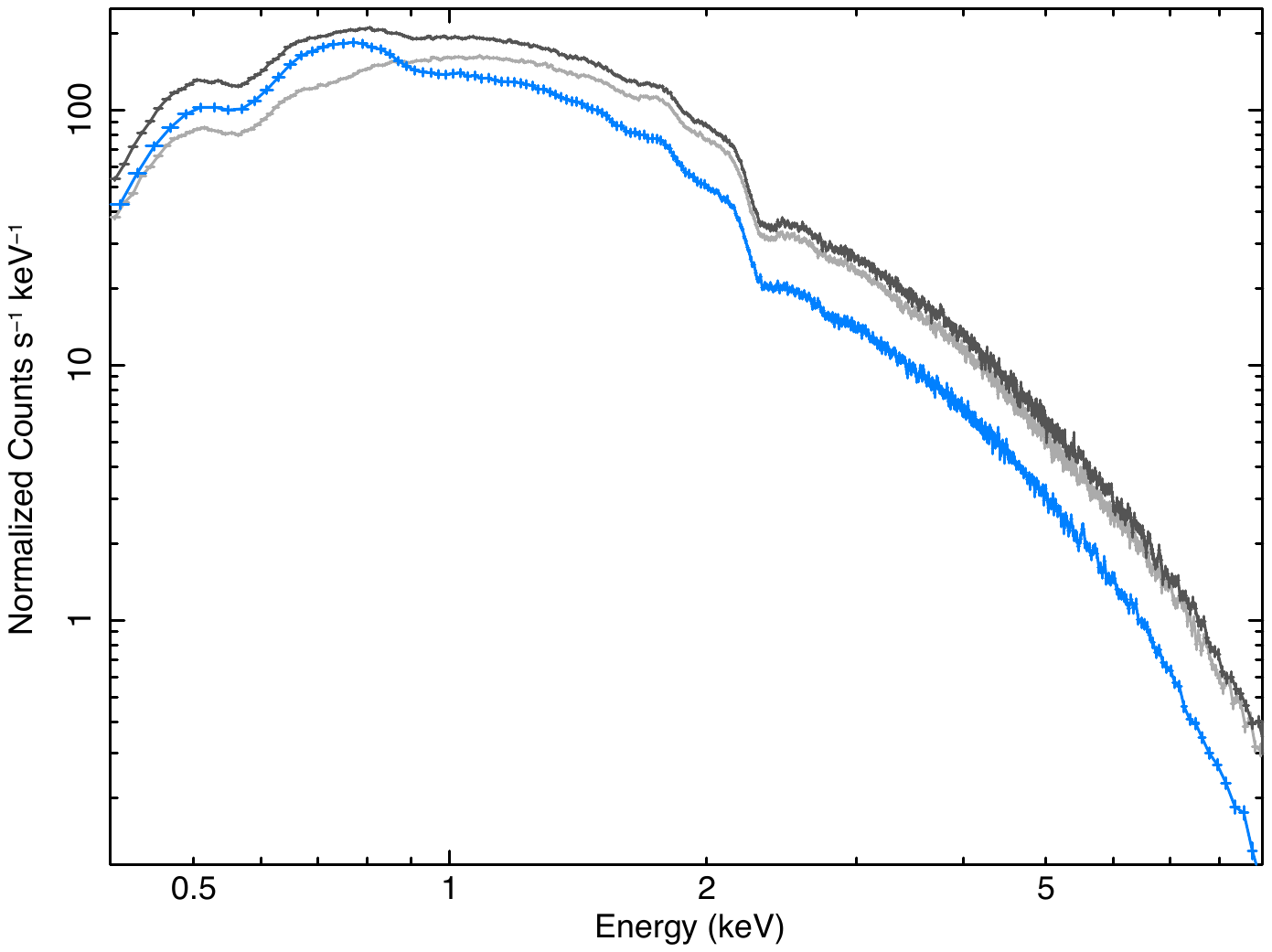}
\caption{Comparison of the $0.4-9$~keV counts spectra for the \nicer observations reported  here (blue) and the observations from 2017 intervals A (light grey) and E (dark grey) from \citet{ludlam19b}. The source is at a lower flux in 2020 in comparison to the previous \nicer observations. }
\label{fig:lcounts}
\end{center}
\end{figure}

\nustar observed \source on 2020 April 19 starting at 07:21:09 UT. ObsID 30601006002 contains $\sim32.3$~ks of data from Focal Plane Module (FPM) A and $\sim32.1$~ks from FPMB.
The \nustar data were reduced using the standard data reduction process with {\sc nustardas} v1.9.2 and {\sc caldb} 20191219.
Spectra and \lc s are extracted using a circular region with a 80$''$ radial centered on the source. Backgrounds were generated from a 80$''$ radial region on the same detector but away from the source. 

There were no Type-I X-ray bursts present in either data set, therefore no further filtering was needed. Systematic errors of 1\% in the $2-10$ keV band and 5\% in the $0.3-2$ keV band were added to the \nicer spectrum \citep{alabarta20}. The \nustar spectra were binned by 3 PI channels using {\sc grppha} \citep{choudhury17}. Figure~\ref{fig:lc} shows the \nustar/FPMA (circles) and \nicer (stars) \lc s binned to 128~s starting from when \nicer began observing \source. The source exhibits $\lesssim10\%$ variability over the $\sim65$ ks of elapsed time since the start of the observations. Using the definitions from \citet{bult18}, we compare the \nicer\ hard color ($3.8-6.8$~keV~/~$2.0-3.8$~keV) and soft color ($1.1-2.0$~keV~/~$0.5-1.1$~keV) of \source to the previous observations that occurred in 2017 August during an enhanced accretion period in Figure~\ref{fig:HID}. The 2020 observations presented here captured the source at a lower intensity. For comparison, we also show the counts spectrum in Figure~\ref{fig:lcounts} for the 2020 \nicer observations to intervals A and E from \citet{ludlam19b}. 

Furthermore, we search the data for pulsations in \S \ref{sec:timing}. The data obtained for both NICER and NuSTAR were barycentered to the solar system barycenter using the source position prior to the search. We used the same source regions as previously mentioned to extract source photons in the $3 - 78$~keV energy band from the \nustar observations. \nicer photons were extracted from the $0.3-10$~keV energy band. Note that \nicer is not an X-ray imaging mission, therefore there is no need for an extraction region. Events were extracted using the same GTIs as were used for extracting spectra. However, the NuSTAR FPMA and FPMB can have GTI mismatches \citep{Bach:2015b}, hence we trimmed each GTI interval to be within a safe range of $100-300$ s. We applied clockfile  {\sc v.108}, generated by using {\sc nustar-clock-utils} \footnote{https://github.com/nustar/nustar-clock-utils}, to the \nustar event files using the FTOOL {\sc barycorr}. The clockfile v.108 corrects for both the NuSTAR clock variations and absolute timing uncertainty of 5~$\mu$s between NuSTAR and NICER.

\begin{figure} 
\begin{center}
\includegraphics[angle=270,width=0.47\textwidth,trim=50 10 40 60,clip]{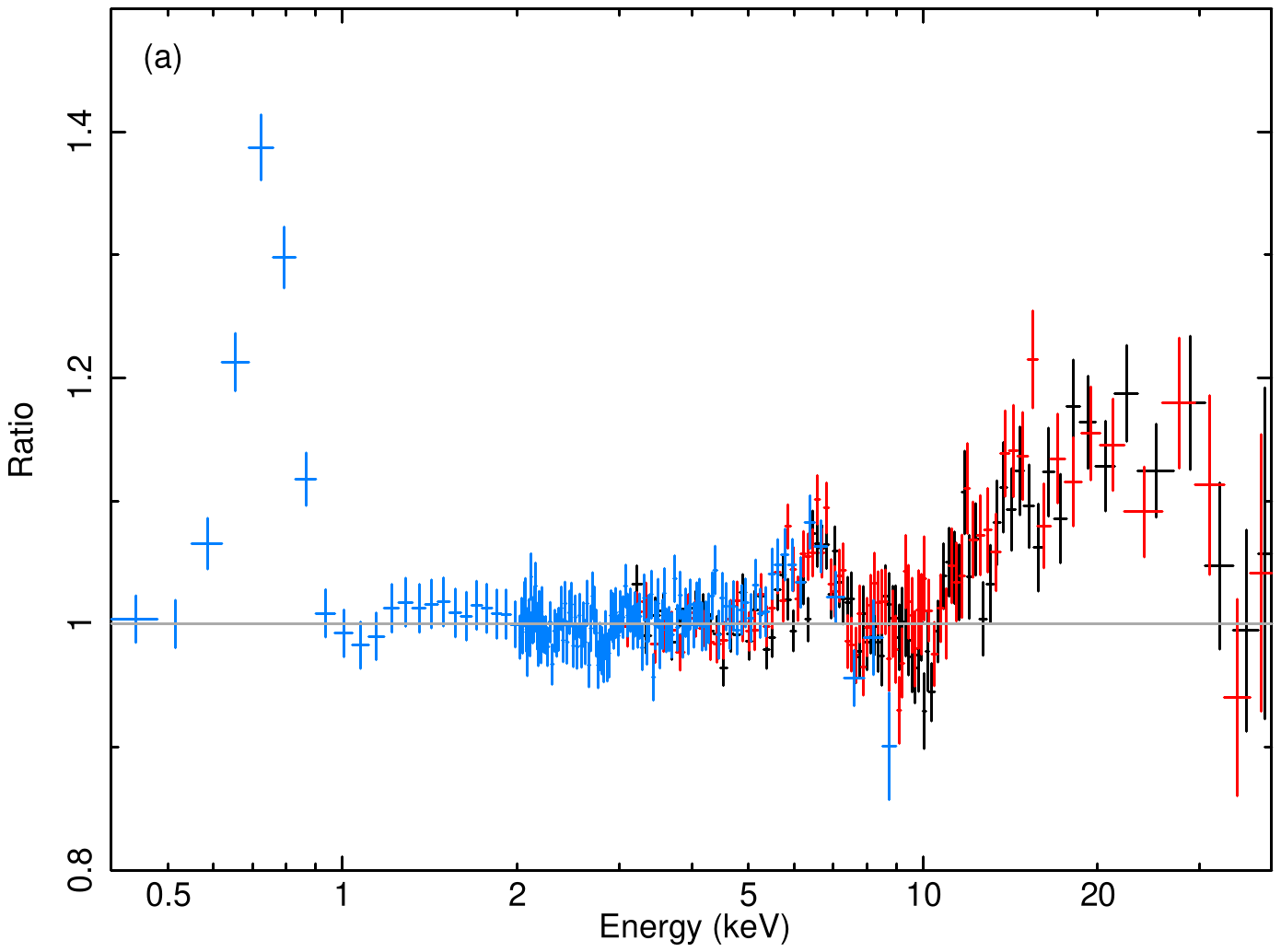}
\includegraphics[angle=270,width=0.47\textwidth,trim=70 10 40 60,clip]{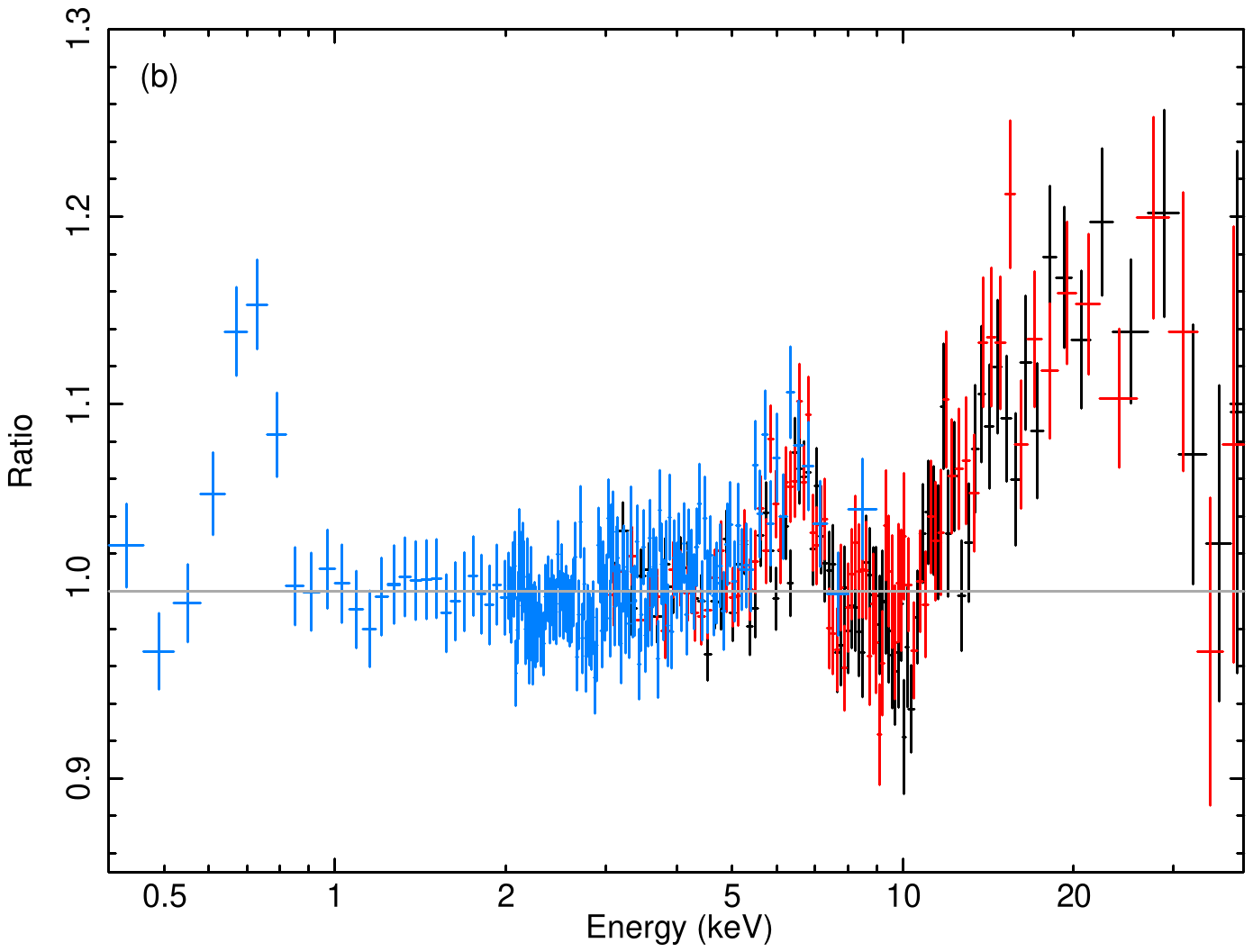}
\caption{Ratio of the \nicer (blue) and \nustar (FPMA: black, FPMB: red) data to the simple continuum model of an absorbed blackbody and power-law (a) without the two edge components and (b) with the edges added. A prominent O emission line is present $\sim0.7$ keV, as well as a Fe K line $\sim6.4$ keV and a Compton Hump at the highest energies. These regions were ignored when fitting the continuum to prevent these features from skewing the fit. Data were rebinned for plotting purposes.}
\label{fig:ratio}
\end{center}
\end{figure}

\section{Analysis and Results} \label{sec:results}
\subsection{Spectral} \label{sec:spec}
The spectral analysis was conducted using {\sc xspec} v.12.11.0 \citep{arnaud96}. The \nicer data were modeled in the $0.4-9$~keV band, whereas the \nustar data were considered in the $3-40$~keV energy range. Data above these energy ranges are dominated by the X-ray background. A constant was allowed to vary for the \nustar/FPMB and \nicer spectra, while the \nustar/FPMA is fixed at 1.0, to allow for cross-calibration differences. The absorption along the line of sight was accounted for with the {\sc tbabs} model \citep{wilms00}. There were two narrow features in the low-energy portion of \nicer\ spectrum that were also seen in \citet{ludlam20} for a different source, 4U~$1735-44$. These are likely astrophysical in origin, i.e., due to the neutral interstellar medium (ISM) along the line of sight \citep{pinto13}, although instrumental uncertainties are also a plausible explanation.
We added two {\sc edge} components with energies bound between $0.5-0.6$~keV and $0.8-0.9$~keV to account for the features.

The continuum was modeled according to the framework of \citet{lin07} in order to provide a direct comparison to the results in the previous analysis on \source using \nicer observations \citep{ludlam19b, koliopanos20b}. A simple absorbed cutoff power-law to account for weak Comptonizaton from the corona and single-temperature thermal component originating from a boundary layer or NS surface were sufficient to describe the continuum spectra. The model parameters and values are reported in Table \ref{tab:refl}. The photon index, $\Gamma$, is softer and the single-temperature blackbody is hotter in comparison to the 2017 observations reported in \citet{ludlam19b} ($\Gamma\leq1.86$, $kT_{bb}<0.83$~keV). However, the thermal component is cooler than the value reported in \citet{koliopanos20b} ($kT_{bb}\sim1.8$~keV). The data do not require a disk component suggesting that the thermal emission from the disk is cooler than when the source was observed in 2017 during an enhanced accretion episode. 

We switch the empirical continuum model for a more physically motivated model. When using {\sc nthcomp} to account for Comptonized accretion instead of the cutoff power-law that would arise from the corona, we find $\Gamma=2.4\pm0.1$, seed photon temperature $kT_{bb}=1.04 _{- 0.03 }^{+ 0.02 } \times10^{-1}$~keV, a high-energy rollover that tends to the upper limit to 1000~keV, and normalization of ${\rm norm}=1.74 _{- 0.01 }^{+ 0.02 }\times10^{-1}$. This is similar in shape to the cutoff power-law component in the previous continuum model description, but predicts photons out to higher-energy that we are not sensitive to with the current data.  We note that the high-enery rollover also tended to 1000~keV when {\sc nthcomp} was applied to the \nicer\ and \integral\ observations of \source \citep{ludlam19b}. The fit still requires a single-temperature blackbody component of $kT=1.43\pm0.01$~keV and ${\rm norm}=1.01 _{- 0.02 }^{+ 0.01 }\times10^{-3}$. The edges and multiplicative constants are similar to the continuum values reported in Table \ref{tab:refl}  with a slightly lower value for $\rm{N_{H}}$ ($3.20 \pm0.02 \times10^{21}$~cm$^{-2}$), but this is likely a more reliable measure of the column density given that the {\sc nthcomp} has a low-energy turn over that the power-law component lacks (which can lead to higher inferred  $\rm{N_{H}}$ value in the latter case). However, the $\rm{N_{H}}$ values between the two continuum models are not largely discrepant.
The reduced $\chi^2$ is upwards of~2.8 ($\chi^2/dof=4236/1464$), but statistically better than the simple continuum model description.
However, there currently does not exist reflection models that are tailored to the atypical abundances observed in the accretion disks of UCXBs using a Comptonized or blackbody illuminating continuum. Therefore, while we report the model parameter values using {\sc nthcomp} for completeness, we do not pursue this further when modeling the reflected emission.

\begin{table*}[t!]
\caption{Joint NICER and NuSTAR Spectral Modeling}
\label{tab:refl} %
%4U1543continuum_cal2020_sys_edges_chur_10272020_1e6_v2.out

%4U1543continuum_cal2020_sys_edges_chur_disklines_diffRin_01192021_1E6.out
%4U1543continuum_cal2020_sys_edges_chur_disklines_sameRin_01192021_1E6.out

%4U1543xillverCO_fit_cal2020_sys_edges_chur_10262020_1e6.out
%4U1543xillverCO_fit_cal2020_sys_edges_noOVIII_chur_lowRi_10262020_1e6.out
%4U1543xillverCO_fit_cal2020_sys_edges_noFe_chur_10262020_1e6.out

\begin{center}
\begin{tabular}{llcccccc}
\hline

Model & Parameter & Continuum & \multicolumn{2}{c}{{\sc diskline}}& \multicolumn{3}{c}{{\sc xillverCO}} \\
&& & D1& D2&X1 & X2 & X3 \\
\hline

{\sc constant} %%%%

& C$_{\rm{FPMB}}$ 
& $ 1.02 \pm 0.01$
& $ 1.01 \pm 0.01$
& $ 1.01 \pm 0.01$
& $ 1.01 \pm 0.01$
&$ 1.01 \pm 0.01$
&$ 1.01 \pm 0.01$
\\

& C$_{\rm{NICER}}$
& $ 1.02 \pm 0.01 $
&$ 1.03 \pm 0.01$
&$ 1.01 \pm 0.01$
&$ 1.03 \pm 0.01$
&$ 1.03 \pm 0.01$
&$ 1.03 \pm 0.01$
\\

{\sc tbabs} %%%%

& $\mathrm{N}_{\mathrm{H}}$ ($10^{21}$ cm$^{-2}$)
& $ 3.35 _{- 0.01 }^{+ 0.02 }$
& $ 3.22 \pm0.01$
&$ 3.20 _{- 0.02 }^{+ 0.01 }$
& $ 3.45\pm0.02$
&$ 3.45^{*}$
&$ 3.36 _{- 0.02 }^{+ 0.03 }$
\\

{\sc edge} %%%%
& E (keV) ($10^{-1}$)
& $ 5.2^{*} $
&$ 5.10 _{- 0.02 }^{+ 0.08 }$
&$ 5.20 \pm0.01$
& $ 5.20 \pm0.01$
&...
&$ 5.13 _{- 0.05 }^{+ 0.04 }$
\\

& $\tau_{\mathrm{max}}$ ($10^{-1}$) 
& $ 3.2\pm0.8 \times 10^{-7}$
&$ 0.657 \pm0.01$
& $ 0.57 _{- 0.02 }^{+ 0.05 }$
& $ 1.15 _{- 0.02 }^{+ 0.03 }$
&...
&$ 1.09 \pm0.04$
 \\
 
{\sc edge} %%%%
& E (keV) ($10^{-1}$)
& $ 8.78 \pm0.02$
&$ 8.88 _{- 0.03 }^{+ 0.06 }$
&$ 8.89 _{- 0.04 }^{+ 0.06 }$
&$ 8.53 _{- 0.04 }^{+ 0.03 }$
&...
&$ 8.54 _{- 0.04 }^{+ 0.03 }$
\\

& $\tau_{\mathrm{max}}$ ($10^{-1}$) 
& $ 3.39\pm0.07$
&$ 1.87 _{- 0.02 }^{+ 0.01 }$
&$ 1.88 _{- 0.08 }^{+ 0.02 }$
& $ 2.61\pm0.03$
&...
&$ 2.37 _{- 0.02 }^{+ 0.03 }$
 \\

{\sc bbody} %%%%
& kT (keV) 
& $ 1.44 \pm 0.01$
&$ 1.31 \pm0.01$
&$ 1.27 \pm0.01$
& $ 1.29 \pm 0.01$
& $ 1.30 _{- 0.01 }^{+ 0.02 }$
&$ 1.29 _{- 0.01 }^{+ 0.03 }$
\\

& norm$_{\mathrm{bb}}$ ($10^{-3}$) 
& $ 1.05 \pm 0.02$
&$ 1.01 \pm 0.01$
&$ 0.98 _{- 0.02 }^{+ 0.03 }$
& $ 1.02\pm0.02$
&$ 1.08 _{- 0.04 }^{+ 0.01 }$
& $ 0.97 _{- 0.01 }^{+ 0.02 }$
\\

& $R_{\mathrm{bb, sph}}$ (km)
& $2.78 \pm 0.06$
& $3.30 \pm 0.04$
&$3.45_{-0.07}^{+0.11}$
& $3.42 \pm 0.07 $
& $3.46 _{-0.13}^{+0.06} $
& $3.33_{-0.04}^{+0.10} $
\\

%& $R_{\mathrm{bb, band}}$ (km)
%& $8.8 \pm 0.2$
%&
%&
%& $10.8 \pm 0.2$
%& $10.94_{-0.4} ^{+0.2}$
%& $ $
%\\

{\sc cutoffpl} %%%%
& $\Gamma$ 
& $ 2.41 \pm 0.01$
& $ 2.34 \pm0.01$
&$ 2.33 \pm0.01$
&$ 2.31 \pm0.01$
&$ 2.34  \pm0.01$
&$ 2.26 \pm0.01$
 \\

& E$_{\rm{cutoff}}$ (keV)
& $ 176\pm8$
& $ 152 \pm2$
&$ 148 \pm3$
&$ 99 _{- 5 }^{+ 7 }$
&$ 137 \pm7$
&$ 74 \pm 1$
\\

& norm$_{\mathrm{pl}}$ ($10^{-1}$) 
& $ 1.82\pm0.01$
&$ 1.64 \pm0.01$
&$ 1.62 _{- 0.02 }^{+ 0.01 }$
&$ 1.30 \pm0.03$
&$ 1.43 _{- 0.07 }^{+ 0.03 }$
&$ 1.09 _{- 0.04 }^{+ 0.02 }$
\\

{\sc diskline}$_{1}$ %%%%
& E$_{\mathrm{O}}$ (10$^{-1}$ keV)
& ...
& $ 6.88 _{- 0.06 }^{+ 0.02 }$
& $ 6.87 _{- 0.02 }^{+ 0.03 }$
& ...
& ...
& ...
\\

& $|q|$
& ...
&$ 2.38 _{- 0.01 }^{+ 0.07 }$
& $2.41 \pm0.02$
& ...
& ...
& ...
\\

& $i$ ($^{\circ}$)
& ...
&$ 52.3 _{- 0.3 }^{+ 1.2 }$
& $ 53 \pm1$
& ...
& ...
& ...
\\

& \rin (\rg)
& ...
&$ 6.02 _{- 0.02 }^{+ 0.20 }$
& $ ^{\dagger}6.01 _{- 0.01 }^{+ 0.20 }$
& ...
& ...
& ...
\\

& \rin (km)
& ...
& $12.44_{-0.04}^{+0.40}$
& $^{\dagger}12.42 _{- 0.02 }^{+ 0.41 }$
& ...
& ...
& ...
\\

& norm$_{\mathrm{line1}}$ (10$^{-2}$)
& ...
&$ 1.61 _{- 0.01 }^{+ 0.05 }$ 
& $ 1.58 \pm0.01$
& ...
& ...
& ...
\\

{\sc diskline}$_{2}$ %%%%
& E$_{\mathrm{Fe}}$ (keV)
& ...
&$ 6.40 _{*}^{+ 0.03 }$
& $ 6.41 _{- 0.01 }^{+ 0.04 }$
& ...
& ...
& ...
\\

& \rin (\rg)
& ...
&$ 12.32 _{- 0.05 }^{+ 0.02 }$
& $ ^{\dagger}6.01 _{- 0.01 }^{+ 0.20 }$
& ...
& ...
& ...
\\

& \rin (km)
& ...
& $25.46_{-0.10}^{+0.04}$
& $^{\dagger}12.42 _{- 0.02 }^{+ 0.41 }$
& ...
& ...
& ...
\\

& norm$_{\mathrm{line2}}$ (10$^{-4}$)
& ...
&$ 3.40 _{- 0.03 }^{+ 0.18 }$
& $ 3.9 _{- 0.1 }^{+ 0.2 }$
& ...
& ...
& ...
\\

{\sc relconv} %%%
& $q$ 
& ...
&...
&...
&$ 2.4\pm0.1$
&$ 2.0 \pm0.2$
&$ 2.55 _{- 0.03 }^{+ 0.02 }$
\\

& $i$ ($^{\circ}$)
& ...
&...
&...
&$ 53 _{- 1 }^{+ 2 }$
&$ 53 _{- 2 }^{+ 1 }$
&$ 53 \pm 1 $
\\

& $R_{\mathrm{in}}$ (\risco)
&...
&...
&...
& $ 1.02 \pm0.01$
&$ 1.03 _{- 0.03 }^{+ 0.04 }$
&$ 1.02 _{- 0.01 }^{+ 0.03 }$
\\

& $R_{\mathrm{in}}$ (\rg)
&...
&...
&...
& $ 6.12 \pm0.06$
&$ 6.18 _{- 0.18 }^{+ 0.24 }$
&$ 6.12 _{- 0.06 }^{+ 0.18 }$
\\

& $R_{\mathrm{in}}$ (km)
&...
&...
&...
& $ 12.7 \pm0.1$
&$ 12.8 _{- 0.4 }^{+ 0.5 }$
&$ 12.7 _{- 0.1 }^{+ 0.4 }$
\\

\xillverco 
& $A_{\rm{CO}}$
&...

&...
&...
& $ 4.3 \pm0.1$
&$ 4.3 _{- 0.1 }^{+ 0.2 }$
&$ 4.0 \pm 0.1 $
\\

& $kT_{\rm{disk}}$ (10$^{-2}$ keV)
& ...
&...
&...
&$ 5.02 _{- 0.01 }^{+ 0.02 }$
&$ 5.01 _{- 0.01 }^{+ 0.03 }$
&$ 5.01 \pm 0.01$
\\

& Frac$_{\rm{PL/BB}}$ (10$^{-1}$)
&...
&...
&...
&$ 1.18 \pm0.02$
&$ 1.15 _{- 0.07 }^{+ 0.09 }$
&$ 1.51 _{- 0.03 }^{+ 0.02 }$
\\

& norm$_{\mathrm{xillver}}$ ($10^{-8}$)
& ...
&...
&...
&$ 2.49\pm0.05$
&$ 2.0 \pm 0.1$
&$ 2.31 _{- 0.03 }^{+ 0.06 }$
\\

\hline
&$\chi^{2}$ (dof) & 4935 (1462)  & 1732 (1453) & 1748 (1454) & 1814 (1454) & 1613 (1408) & 1355 (1101)\\ 
\hline
$^{*}=\rm{fixed}$ & $^{\dagger}=\rm{tied}$
\end{tabular}
\end{center}

\medskip
Note.---  Errors are reported at the 90\% confidence level and calculated from Markov Chain Monte Carlo (MCMC) of chain length $10^{6}$. \nicer\ is fit in the $0.4-9$ keV energy band while \nustar is fit in the $3-40$ keV band.  A multiplicative constant is used on the \nicer and FPMB data, while FPMA is fixed to unity. The spherical blackbody radius is calculated assuming a distance of 7~kpc and color correction factor of 1.7 \citep{shimura95}. The emissivity index and inclination are tied between the two {\sc diskline} components. D1 allows the inner disk radii to differ between the two {\sc diskline} components, whereas D2 assumes a common emission radius for both lines. The outer disk radius is fixed at 990 \rg and the dimensionless spin parameter is set to $a_{*}=0$ (hence, 1 \risco = 6 \rg = 12.4~ km). The photon index and high-energy cutoff in the \xillverco model are tied to the values of the continuum power-law component. X1 is the full passband from $0.4-40$~keV,  X2 is ignoring the the \ion{O}{8} line by ignoring below $0.9$ keV and fixing the column density and low-E edges, and X3 uses the $0.4-40$~keV band but ignores the Fe line region from $5-8$~keV. \\ \\
\end{table*}

Fitting the emission lines with simple Gaussian components provides an equivalent width of $\sim39$~eV for the \ion{O}{8} near 0.7~keV and $\sim143$~eV for the Fe~K emission at 6.4~keV. These are consistent with the values reported in \citet{ludlam19b}. 
For direct comparison to \citet{ludlam19b}, we add two {\sc diskline} \citep{fabian89} components to account for the \ion{O}{8} and Fe~K lines with energies between $0.6-0.7$~keV and $6.4-6.97$~keV, respectively. The inclination ($i$) and emissivity index ($|q|$) parameters are tied between the line components. In the first instance, we allow the inner disk radius to differ between components. This is reported in Table \ref{tab:refl} under D1. The emitting radius of the Fe line is further out in the disk than the \ion{O}{8} line, which is consistent with the results reported in \citet{ludlam19b} when \rin is allowed to differ.  In the second case, we tie the inner disk radius between the two lines, which is reported under D2 in Table \ref{tab:refl}. In this case, the emission region of both lines is from the inner most accretion disk within $\leq6.21$~\rg, which agrees with the inferred inner disk radius from interval E ($<8.7$~\rg, \citealt{ludlam19b}) and suggests that the disk has not receded after the peak flux observed in 2017. For both spectral fits, the emissivity index is consistent with values reported in \citet{madej14} and observed in other NS LMXBs such as 4U~1705$-$44, 4U~1636$-$53, 4U~1702$-$429, Serpens~X-1, as well as the UCXB 4U~0614+091 \citep{egron13, ludlam17a, ludlam18,ludlam19a}.  Additionally, the inclination is lower ($i\sim53^{\circ}$) than has been reported previously for this source, but in agreement with the inclination inferred from the optical observation \citep{wang04}. 

It is important to note that while using {\sc diskline} is acceptable as a preliminary diagnostic for line emission, the profile assumes a single emission line is being broadened by Doppler and relativistic effects. This does not account for the blending of emission from other atomic species or energy levels within the energy region of interest for the emission lines (e.g., the blending of \ion{Fe}{25} and \ion{Fe}{26} K$_{\alpha}$,  \ion{O}{8} Lyman $\alpha$ and $\beta$, or even emission of Mg blended with Fe~L as shown in \citealt{ludlam18}). The full reflection spectrum is a series of atomic features that are superimposed onto a reprocessed continuum that is then broadened. Hence, a complete reflection model should be utilized when performing spectral modeling of reflection. 

We opt for a more consistent approach to describe the reflection spectrum present within the system by using a modified version of {\sc xillver} \citep{garcia13} that accounts for the unusual elemental abundances in the accretion disk, known as \xillverco. This assumes that coronal emission is illuminating the accretion disk as a power-law, $\Gamma$, with a high-energy cutoff, E$_{\rm{cutoff}}$. The reflection model also contains emergent thermal emission from the accretion disk itself, $kT_{\rm{disk}}$, at the location where the emission features arise. The $\rm{Frac}$ parameter adjusts the strength of the the power-law illuminating the disk relative to blackbody arising from the disk ($\sigma T^{4}$), $\rm{Frac=Flux_{PL}(10^{2}-10^{6}\ eV)/Flux_{BB}(0.1-10^6\ eV)}$. This is an updated version of the model used in \citet{madej14}. The earlier grid calculations had set the abundances of all elements to be 10 times those from \citet{lodders03}, except for H and He (which were left at solar abundance), and in the case of C and O to 100 times the abundance from \citet{lodders03}. 
Here, the updated table of the \xillverco model has the abundances set as follows: H and He to 0.1 times solar abundance from \citet{lodders03}, C and O are allowed to vary using the $A_{\mathrm{CO}}$ parameter, and all other elements are set to solar abundance. This model also has over $10^{5}$ more spectral grid points than the initial model used in \citet{madej14}. 

When using \xillverco, we tie the photon index and high-energy cutoff to those in the continuum power-law component for consistency. The reflection component is convolved with {\sc relconv} \citep{dauser10} to account for broadening due to different effects within the innermost region of the accretion and proximity to the NS. We tie the inner and outer emissivity index in order to create a single illumination profile, $q$. The outer disk radius is set to $990$~\rg and the dimensionless spin parameter is fixed at $a_{*}=0$.

\begin{figure} 
\begin{center}
\includegraphics[angle=270,width=0.47\textwidth,trim=20 0 20 70,clip]{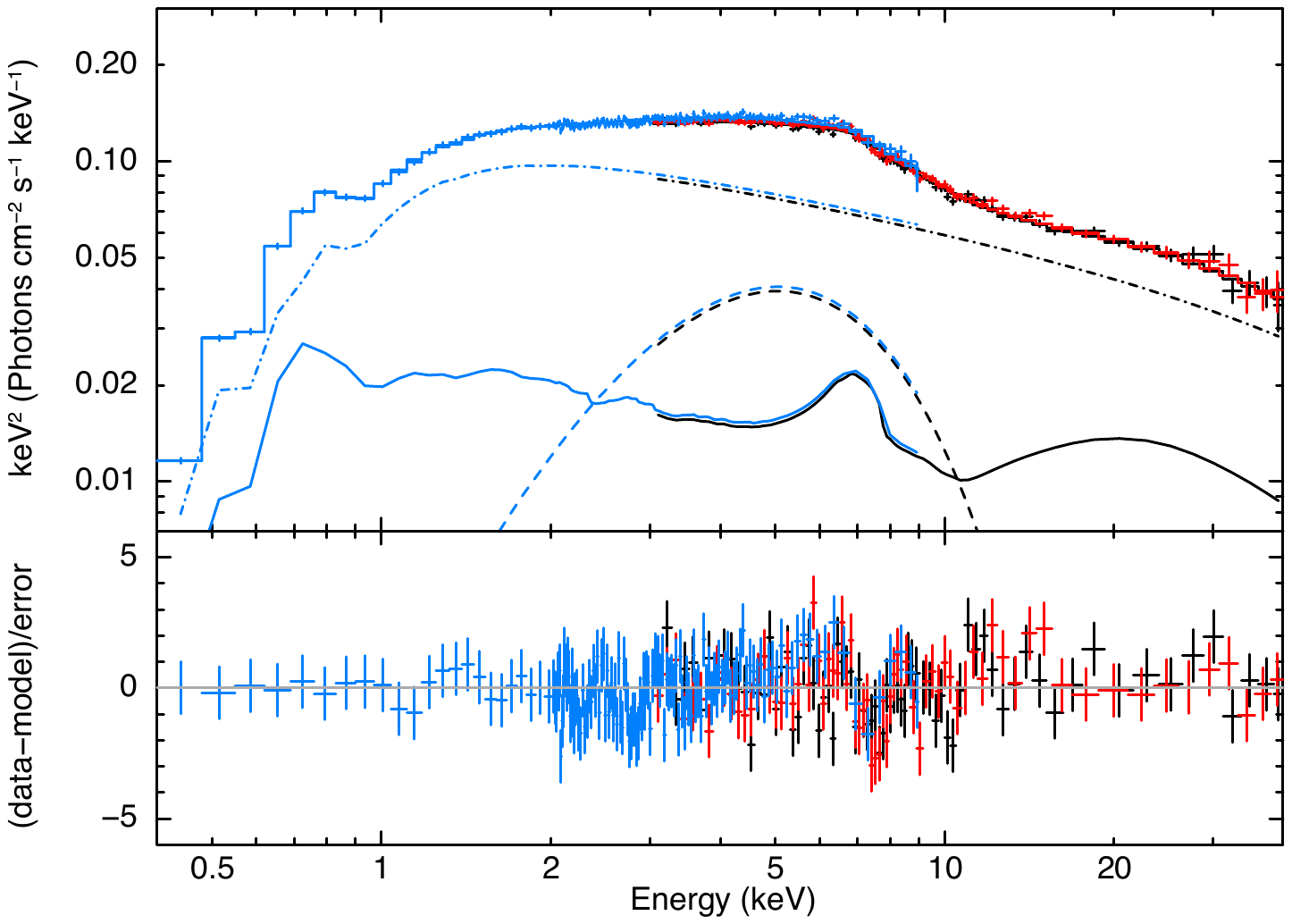}
\caption{The unfolded spectrum with model components for the \nicer (blue) and \nustar (FPMA: black, FPMB: red) data for the reflection modeling reported in Table~\ref{tab:refl}. The dashed line indicates the single-temperature blackbody, the dot-dashed line is the power-law component, the solid line is the reflection component from \xillverco. }
\label{fig:model}
\end{center}
\end{figure}

Applying this model to the full $0.4-40$~keV, we achieve an improved fit of $\Delta \chi^{2} = 3121$ for 8 degrees of freedom (dof) in comparison to the continuum fit. The values for each parameter are shown in Table~\ref{tab:refl} under  X1. 
While the statistical fit may be worse than the overall model using {\sc diskline} components, using \xillverco provides more information regarding the emitting material (e.g., abundance of C/O) and correctly accounts for the reprocessed continuum emission.
The unfolded model and residuals divided by the error are shown in  Figure~\ref{fig:model}. 
Note that the discrepancy above 6 keV between \nicer and \nustar data was also reported in  \citet{ludlam20} when fitting simultaneous data from both missions. This has to do with the difference in calibration between missions (see \citealt{ludlam20} for a more detailed discussion).

\begin{figure} 
\begin{center}
\includegraphics[width=0.47\textwidth,trim=0 0 0 0,clip]{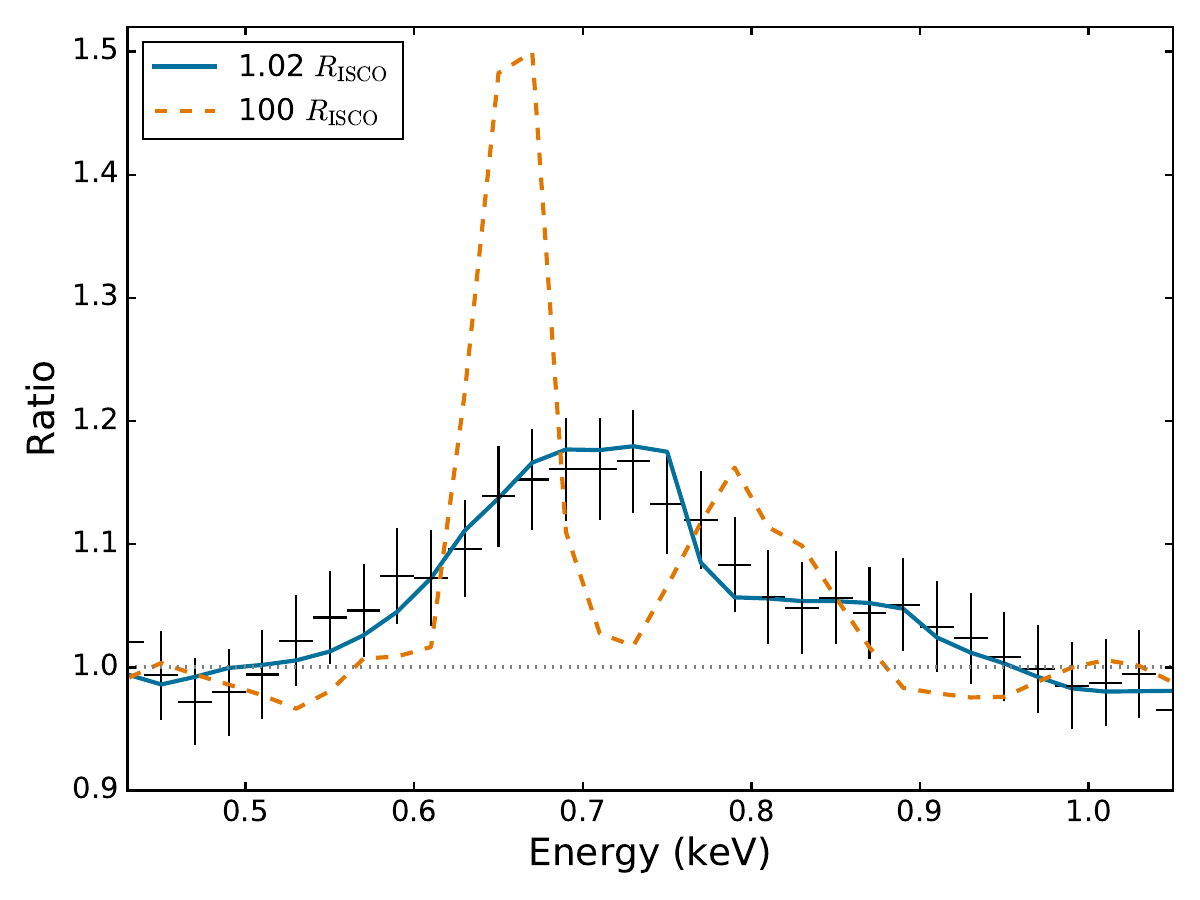}
\caption{Best-fit reflection model reported in Table~\ref{tab:refl} at 1.02 \risco (solid line) and contrasting 100 \risco (dashed line) overlaid on the NICER data to highlight the broad O line component. The larger inner disk radius relaxes the relativistic effects to show the local rest-frame emission. The \ion{O}{8} Lyman $\alpha$ and $\beta$ components can be seen when relativistic effects are removed. For clarity, the data were rebinned.
}
\label{fig:Oline}
\end{center}
\end{figure}

The disk is close to the inner most stable circular orbit (\rin$ = 1.02\pm0.01$~\risco) and the inclination is consistent with the values inferred from the {\sc diskline} modeling ($i\sim53^{\circ}$). 
Figure~\ref{fig:Oline} shows the \ion{O}{8} line with the blurred reflection model at the best fit inner disk radius overlaid. For reference, we have also plotted the reflection model at a large radius to remove the relativistic effects. The \ion{O}{8} Lyman $\alpha$ and $\beta$ components become evident when relativistic effects are relaxed. The abundance of C/O is about ten times less than the values reported by \citet{madej14} when using the previous version of \xillverco, but it is important to note that the abundances in that model were set up ten times larger. Therefore the values obtained for the C/O abundance are consistent.  
The Frac parameter is in agreement with the value reported in \citet{madej14}.
The thermal emission from the accretion disk is indeed cooler ($\sim0.05$~keV) than during the 2017 \nicer\ observations at peak intensity ($\sim0.1$~keV: \citealt{ludlam19b}). 

To check if this lower disk temperature is consistent with not being able to detect the accretion disk component in the continuum modeling, we add a {\sc diskbb} component to the continuum description with $kT=0.05$~keV and the normalization value equivalent to the inner edge of the accretion disk inferred from reflection modeling (norm$_{disk}=26$ for $i=53^{\circ}$, $D=7$~kpc, and a color-correction factor of $1.7$). The {\sc diskbb} component accounts for less than 0.00001\% of the photons at 0.5 keV, which is consistent with not being statistically needed during the simple continuum modeling. We can also calculate the expected thermal flux from the \xillverco model itself that would be expected for a distant observer. The expected unabsorbed thermal flux at 7~kpc in the $0.1-10^{6}$ eV band would be ${\rm F_{xillverCO,\  BB,\ 7~kpc}}=2.11\times10^{-14}$~\fluxcgs. For comparison, the unabsorbed continuum flux from the source in the same energy band using `energies extend' command in {\sc xspec} is ${\rm F_{continuum,\ 0.1-10^{6}\ eV}}\simeq 3.53\times10^{-8}$~\fluxcgs. This is the same order of magnitude contribution as the check using {\sc diskbb} and consistent with the disk component not being detected in the overall continuum model.

To check if these line components originate from a concurrent radius in the accretion disk or different radii, we model the spectra by fixing the absorption column and removing the \ion{O}{8} line by ignoring below 0.9~keV so that the fit will be driven by the Fe line (X2 in Table~\ref{tab:refl}). This provides a position on the inner disk radius of \rin$ = 1.03_{-0.03}^{+0.04}$~\risco from fitting the reflection emission without the O line. 
Conversely, we also fit the spectrum from $0.4-40$~keV but ignore the Fe band from $5-8$~keV to see what constraints are returned from the O line (X3 in Table~\ref{tab:refl}). This gives an inner disk of \rin$ = 1.02_{-0.01}^{+0.03}$~\risco. The emission radii inferred from each line are consistent within the 90\% confidence level, supporting a common emission radius in this system as was suggested from the line profiles plotted in velocity space (see Fig.~5 of \citealt{ludlam19b}). 

\subsection{Timing} \label{sec:timing}
We searched the \nicer\ and \nustar\ data for coherent pulsations that would provide further support for a NS accretor in this system. We utilized the {\sc HENDRICS} package within the powerful Python X-ray timing software {\sc Stingray} \citep{Hup:2019} to search for pulsations. Light curves binned with 0.001~s, 1~s, and 10~s were generated via {\sc HENcalibrate} and {\sc HENlcurve} to check for any variations or non-stationary processes within the \lc, but none were found. We proceeded to split the (unbinned) time-series events into chunks of 1/10th the orbital period in order to conduct a search for pulsations. 
In the case of highly compact binaries, Doppler effects due to orbital motion varies the pulse frequency and spreads power over multiple frequency bins. 
To mitigate this, a constant acceleration model was used to account for the orbital motion (as outlined in \citealt{Ransom,Ransom:2002}) and searches were conducted over various possible acceleration values. 
The maximum acceleration value for these searches is given by \mbox{$Acc=\frac{z~c}{f~T^2}$}, where $z$ is the acceleration search depth \mbox{($z = \dot{f} \times T^2$)}, $c$ is the speed of light, $f$ is the pulse frequency, and $T$ is the observation duration. 
We therefore split the time-series in 150~s chunks and then, assuming a bin depth of $z=10$ and maximal pulse frequency of 800~Hz (which would be more rapid than the fastest known milli-second pulsar, PSR~J1748$-$2446ad:  \citealt{Hessels}), we conducted the constant acceleration search. 
The acceleration searches do not yield any significant candidate for coherent pulsations.

\section{Discussion} \label{sec:discussion}
We present the first \nustar observation of the ultra-compact X-ray binary \source that was coordinated with \nicer. The source was in a lower flux state than previously observed by \nicer during the 2017 enhanced accretion phase. The $0.5-50$~keV luminosity was $L_{0.5-50\ \mathrm{keV}} = 4.9 (D/7\ \mathrm{kpc})^{2}\times10^{36}$~\lumcgs, which is 42\% of the peak luminosity during the 2017 brightening.
At this luminosity, the source has a mass accretion rate of \mdot$\ =4.3\times10^{-10}$~\ms year$^{-1}$. The source exhibited strong emission features due to the reprocessing of direct continuum emission by the accretion disk. Fitting the reflection spectrum with a model tailored to the atypical abundances found in these systems, \xillverco, we test the emission radii of the \ion{O}{8} and Fe line components. We find a common emission radius for both line features of \rin$ < 1.07$~\risco, indicating that the disk remains close to the compact object.  There are other systems where the accretion disk is consistent with the innermost stable circular orbit at different flux levels and spectral states (e.g., 1RXS~J180408.9$-$34205: \citealt{ludlam16, degenaar16}), so while \source is not unique in this regard, it is interesting that the disk has not receded after the peak intensity observed in 2017.

Given the amount of evidence supporting a NS accretor in this system (e.g., tentative association with a Type-I X-ray burst: \citealt{serino18}; X-ray--Radio luminosity: \citealt{ludlam19b}), we discuss the results of the spectral modeling in the context of the source being a NS. 
The measured position of the inner disk radius from \xillverco corresponds to $12.4-13.3$~km when assuming a canonical NS mass of 1.4~\ms. The upper limit on the inner disk position (\rin$ = 6.42$~\rg) and $0.5-50$~keV unabsorbed flux of $F_{\mathrm{unabs}}=8.4\times10^{-10}$~\fluxcgs places an upper limit on the dipolar magnetic field of $B\leq0.7(D/7\ \mathrm {kpc})\times10^{8}$~G at the poles. This is within the range estimated from the 2017 enhanced accretion event and further supports a weak B-field in this system. Additionally, this is consistent with the range of magnetic field strengths estimated in \citet{mukherjee15} for accreting millisecond X-ray pulsars (AMXPs), though no pulsations have been detected for this system.

Given the lack of pulsations detected and the small inner disk radius inferred from the reflection features, it does not appear that material is being channeled along magnetic field lines onto the surface of the NS, but rather that the accreting material forms a boundary layer region between the disk and NS surface. 
This would correspond to the single-temperature thermal component in the spectral modeling. The normalization of the blackbody component suggests a compact emission region of 3.4~km at 7 kpc and using a color correction factor of 1.7 \citep{shimura95}. However, this conversion assumes spherical emission rather than banded emission from the NS surface. Accounting for a narrow equatorial banded region with a vertical height that is 5\%$-$10\% of the radius \citep{PS01} can easily increase this blackbody emission radius to $R_{\mathrm{BB}}\sim 11-15$~km. Equation 25 from \citet{PS01} allows us to estimate the size of the boundary layer region normal to the NS surface given the mass accretion rate of \mdot$\ =4.3\times10^{-10}$~\ms year$^{-1}$ at the time of the observation. This corresponds to a boundary layer with a radial extent of $R_{\mathrm{BL}}=1.2$~km from the NS surface. This estimate combined with the position of the inner edge of the accretion disk places an upper limit on the radial extent of the NS of  $R_{\mathrm{NS}} \leq 12.1$~km for $M_{\mathrm{NS}}=1.4$~\ms, if indeed the compact object is a NS in this system.

\citet{koliopanos20b} recently looked at the presence of the Fe line feature in \source over a 20 year period from \rxte (1997), \chandra (2000), \xmm (2001), and \nicer (2017). The source was in a soft state for all the observations that were analyzed with a $0.5-30$~keV luminosity (at 7~kpc) ranging from  $5.1\times10^{36}$~\lumcgs to $6.7\times10^{36}$~\lumcgs. The Fe line was clearly present in the \rxte and \nicer data, but not detected in the \xmm or \chandra observations. Given the small range in luminosity and the consistency of spectral parameter values obtained across missions, the disappearance of the Fe line component is attributed to microscopic processes in the disk rather than macroscopic changes \citep{koliopanos20b}. Though the observations presented here occurred at a lower $0.5-30$~keV luminosity of $\sim4.76\ (D/7\ \mathrm {kpc})^{2}\times10^{36}$~\lumcgs, we see a strong Fe line component. 
The concurrent strong O line indicates that the oxygen in the disk is not fully ionized but rather only partially ionized, yet we do not see screening effects that are predicted to quench the Fe line as per \citet{koliopanos13}.

Though the ionization is not explicitly returned as a parameter by \xillverco model, we can estimate the ionization state of the emitting material via $\xi=4\pi F_{x}/n$ (as is defined for all \xillver models, \citealt{garcia13}), where $F_{x}$ is the ionizing flux from $0.1-1000$~keV and $n$ is the number density of the material in the disk. The \xillverco model has a hard-coded disk number density of $n=10^{17}$~cm$^{-3}$ \citep{madej14}. Using this hard-coded disk number density and that $F_{x}={\rm Flux_{PL}}={\rm Frac}(\sigma T^{4})$ by model definition, then $\log(\xi/[\mathrm{erg\ cm\ s^{-1}}])\simeq1.9-2.1$. This is in line with the emitting material being partially ionized rather than fully ionized, but should be considered a lower limit on the ionization state of the material given that illuminating blackbody X-rays from the boundary layer are not included in the model definition of $\xi$.

Ideally, when modeling the reflection emission in these systems, we would like to be able to account for illumination from the boundary layer or NS surface in addition to the coronal emission. We are currently working to expand the \xillverco model to account for irradiation of the accretion disk by both components, tracking the ionization, and higher disk density. However, these initial results with the current \xillverco model with additional grid points and a more realistic handling of the atypical abundance in these systems than the preliminary grid used in \citet{madej14}, demonstrates the utility of reflection modeling to determine the emergent radius of multiple reflection features. It is unclear if the O and Fe line arising from a common emission radius within the accretion disk is unique to \source or not, but through observing more UCXBs with \nicer and \nustar, as well as with future X-ray missions like Athena \citep{nandra13}, HEX-P \citep{harrison18}, and STROBE-X \citep{ray18}, we can ascertain the accretion geometries of these systems. 
\\

{\it Acknowledgements:} Support for this work was provided by NASA through the NASA Hubble Fellowship grant \#HST-HF2-51440.001 awarded by the Space Telescope Science Institute, which is operated by the Association of Universities for Research in Astronomy, Incorporated, under NASA contract NAS5-26555. Additional support was provided from NASA under the NICER Guest Observer grant 80NSSC21K0121. This research has made use of the NuSTAR Data Analysis Software (NuSTARDAS) jointly developed by the ASI Science Data Center (ASDC, Italy) and the California Institute of Technology (Caltech, USA). J.A.G. acknowledges support from NASA grant 80NSSC20K1238 and from the Alexander von Humboldt Foundation. E.M.C. gratefully acknowledges support through NSF CAREER award number AST-1351222. J.A.T. acknowledges partial support from NASA under NICER Guest Observer grant 80NSSC19K1445. D.J.K.B. acknowledges funding from the Royal Society.

\facilities{ADS, HEASARC, \nicer, \nustar}
\software{HEAsoft (v6.27.2, \citealt{heasarc14}), nustardas (v1.9.2), nicerdas (2020-04-23\_V007a), xspec (v12.11.0, \citealt{arnaud96}), Stingray \citep{Hup:2019}}

\end{document}